# Technoeconomic Supplement of P2G Clusters with Hydrogen Pipeline for Coordinated Renewable Energy and HVDC Systems

Jiarong Li, *Student Member, IEEE*, Jin Lin, *Member, IEEE*, Yonghua Song, *Fellow, IEEE*, Jinyu Xiao, Feng Liu, *Senior Member, IEEE*, Yuxuan Zhao, Sen Zhan

***Abstract*—Under the downward tendency of prices of renewable energy generators and upward trend of hydrogen demand, this paper studies the technoeconomic supplement of P2G clusters with hydrogen pipeline for HVDC to jointly consume renewable energy. First, the planning and operation constraints of large-capacity P2G clusters is established. On this basis, the multistage coordinated planning model of renewable energy, HVDCs, P2Gs and hydrogen pipelines is proposed considering both variability and uncertainty, rendering a distributionally robust chance-constrained (DRCC) program. Then this model is applied in the case study based on the real Inner Mongolia-Shandong system. Compared with energy transmission via HVDC only, P2G can provide operation supplement with its operational flexibility and long term economic supplement with increasing demand in high-valued transportation sector, which stimulates an extra 24 GW renewable energy exploration. Sensitivity analysis for both technical and economic factors further verifies the advantages of P2G in the presence of high variability due to renewable energy and downward tendency of prices of renewable energy generators. However, since the additional levelized cost of the P2G (0.04 RMB/kWh) is approximately twice the HVDC (0.02 RMB/kWh), P2G is more sensitive to uncertainty from both renewable energy and hydrogen demand.**

***Index Terms*— P2G; HVDC; coordinated planning; DRCC**

## NOMENCLATURE

### A. Indicators

- $i/j$ — Indicator of source/demand region
- $ij/l$ — Indicator of HVDC transmission lines and hydrogen pipelines
- $k$ — Indicator of P2G farms
- $t$ — Indicator of time periods
- $s$ — Indicator of scenarios
- $y/\tau$ — Indicator of planning epoch
- M — Indicator of P2G technologies
- U — Indicator of hydrogen downstream sectors

### B. Variables

*a. Planning Variables*

- $P_{i,y}^{\mathrm{WT}} / P_{i,y}^{\mathrm{PV}}$ — Planning capacity of wind turbines/ photovoltaics
- $\sigma_{ij,y,l}$ — Planning status of HVDC transmission line and hydrogen pipeline
- $\delta_{i,y,k}^{\mathrm{M}} / \chi_{i,y}^{\mathrm{M}}$ — Planning status and online number of P2G farms

*b. Operation Variables*

- $P_{i,y,s,t}^{\mathrm{WT,E}} / P_{i,y,s,t}^{\mathrm{WT,H}}$ — Power for electricity transmission/hydrogen production of wind turbines
- $P_{i,y,s,t}^{\mathrm{PV,E}} / P_{i,y,s,t}^{\mathrm{PV,H}}$ — Power for electricity transmission/hydrogen production of photovoltaics
- $f_{ij,y,s,t}^{\mathrm{HVDC}}$ — Power flow of HVDC transmission lines
- $P_{j,y,s}^{\mathrm{D}}$ — Multiplier of unit electric demand profile
- $m_{ij,y,s,t}^{\mathrm{HP}}$ — Hydrogen quantity stored in hydrogen pipelines
- $P_{i,y,s,t}^{\mathrm{M}}$ — Power of P2G cluster
- $P_{i,y,s,t}^{\mathrm{M,ON}}$ — Power of single P2G facility in ON status
- $m_{i,y,s,t}^{\mathrm{M}}$ — Hydrogen production of P2G cluster
- $x_{i,y,s,t}^{\mathrm{M}}$ — ON/OFF status of P2G facility or the sum number of facilities in ON status in the cluster
- $u_{i,y,s,t}^{\mathrm{M}}$ — BOOTING status of P2G facility or the sum number of facilities in BOOTING status in the cluster
- $m_{i,y,s,t}^{\mathrm{S,U}}$ — Hydrogen production rate for source region
- $m_{i,y,s,t}^{\mathrm{D,U}}$ — Hydrogen production rate for demand region as well as the input rate of HP in source side
- $m_{j,y,s,t}^{\mathrm{D,U}}$ — Hydrogen output rate of HP in demand side

### C. Parameters and Sets

- $P_{i,y}^{\mathrm{WT,max}} / P_{i}^{\mathrm{WT,max}}$ — Maximum capacity of new wind turbines (8 GW/40 GW)
- $P_{i,y}^{\mathrm{PV,max}} / P_{i}^{\mathrm{PV,max}}$ — Maximum capacity of new photovoltaics (8 GW/40 GW)
- $P_{i,y}^{\max} / P_{i}^{\max}$ — Maximum capacity of new renewable generators (10 GW/50 GW)
- $c_{y}^{\mathrm{WT}} / c_{y}^{\mathrm{PV}}$ — Investment cost of wind turbines/photovoltaics
- $L_{ij,y}^{\mathrm{HVDC}} / L_{ij}^{\mathrm{HVDC,max}}$ — Maximum number of HVDC transmission lines (1/6)
- $c_{ij}^{\mathrm{HVDC}}$ — Investment cost of single HVDC transmission line including terminals (13.94 billion RMB[1])
- $L_{ij,y}^{\mathrm{HP}} / L_{ij}^{\mathrm{HP,max}}$ — Maximum number of hydrogen pipelines (1/4)
- $c_{ij}^{\mathrm{HP}}$ — Investment cost of single hydrogen pipeline including compressors (32.81 billion RMB)
- $K_{i,y}^{\mathrm{M}} / K_{i,y}^{\mathrm{M,max}}$ — Maximum number of new/online P2G farms (10/50)
- $c_{y}^{\mathrm{M}}$ — Investment cost of single P2G farm
- $N^{\mathrm{M}}$ — Number of facilities in one P2G farm (100)
- $p_{i,s,t}^{\mathrm{WT}} / p_{i,s,t}^{\mathrm{PV}}$ — Prediction mean value of profiles of wind turbines/photovoltaics (p.u.)
- $f_{ij}^{\mathrm{HVDC,max}}$ — Maximum capacity of single HVDC transmission line (8GW)
- $m_{ij}^{\mathrm{HP}}$ — Maximum input rate of single hydrogen pipeline (390 t/h)

[1] 1RMB≈0.1525US$ (according to the exchange rate on 27th, Dec. 2020)

| | |
|---|---|
| $m_{ij}^{\mathrm{HP,max}}$ | Maximum storage capacity of single hydrogen pipeline (12 kt) |
| $p_{j,s,t}^{\mathrm{D}}$ | Prediction mean value of profiles of electric load (p.u.) |
| $P^{\mathrm{M,min}}$ / $P^{\mathrm{M,max}}$ | Minimum/maximum power of single P2G facility in ON status (2/10 MW) |
| $\Delta P^{\mathrm{M,max}}$ | Maximum ramping power of single P2G facility in ON status (10 MW/h) |
| $P^{\mathrm{M,boot}}$ | Booting power of single P2G facility in BOOTING status (1.5 MW) |
| $c^{\mathrm{M,boot}}$ | Booting cost of single P2G facility (500 RMB) |
| $Y^{\mathrm{M}}$ | Lifetime of P2G farm (10 years) |
| $m_{i,y}^{\mathrm{S,U}}$ / $m_{j,y}^{\mathrm{D,U}}$ | Maximum hydrogen demand in source/demand region |
| $c_{y}^{\mathrm{E}}$ | Electricity price |
| $c_{y}^{\mathrm{S,U}}$ / $c_{y}^{\mathrm{D,U}}$ | Hydrogen price of source/demand sectors |
| $k^{\mathrm{WT}}$ / $k^{\mathrm{PV}}$ / $k^{\mathrm{HVDC}}$ / $k^{\mathrm{HP}}$ / $k^{\mathrm{M}}$ | Ratio of fixed operation cost to capacity cost of wind turbines/photovoltaics/HVDC transmission lines/hydrogen pipelines/P2Gs (2%/2%/5%/5%/3%) |
| $Y$ | Number of planning epochs (6) |
| epoch | Years of one planning epoch (5) |
| $S$ | Number of scenarios (4) |
| $D_s$ | Days of scenario $s$ (78/59/155/73) |
| $T$ | Hours of intraday periods (24) |
| $\Omega^{\mathrm{HVDC}}$ | Set of HVDC transmission lines |
| $\Omega^{\mathrm{HP}}$ | Set of hydrogen pipelines |

# I. INTRODUCTION

## 1. Background and Motivation

The sustainable exploration and utilization of renewable energy has been a worldwide trend. Due to the worldwide situation that there is a spatial discrepancy between energy sources and demand such as U. S. and China [1], [2], high-voltage direct current (HVDC) transmission lines are commonly deployed for a long-distance electricity delivery [3]. However, this mode would face both technical and economic issues [4]: 1) with the penetration of volatile renewable energy (RE) increases, more and more flexible resources are required for HVDC transmission to match the source and demand profiles and ensure its utilization rate; 2) with a sharp decreasing trend in the investment cost of wind and solar facilities in the future, the gap on unit electricity production cost between source and demand regions is narrowing, which means the economy of electricity transmission via HVDC is getting worse.

P2G (power-to-gas) technology is another promising method to consume a large amount of renewable energy. The core of P2G is the energy conversion from electricity to hydrogen, and then hydrogen can be applied in chemical, transportation and heating industries. Substituting gray hydrogen from fossil fuels with green hydrogen from renewable energy-based sources, P2G can also help the decarbonization in downstream sectors of hydrogen. Compared to HVDC facilities: 1) P2G facilities are flexible resources that can cooperate with HVDC to follow the variability of renewable energy, furthermore, followed hydrogen pipeline (HP) can also provide enough buffer; 2) the small capacity and short lifetime of P2G facilities can reduce investment risks and respond to price changes more rapidly. The above advantages have been verified by many worldwide research [5-7] and demonstration projects [8]. Therefore, the combination of HVDC and P2G is a feasible solution for future renewable energy utilization and energy system decarbonization.

Several coordinated studies have been performed on HVDC and P2G. The process should begin with the utilization of offshore wind energy, and an economic model has been established to calculate the cost of both technologies [9], [10]. [11] considers the expansion of both transmission networks and P2Gs; however, this expansion is determined in different planning stages with carbon-oriented objectives, which makes it difficult to reflect the technoeconomic supplement of P2G.

In this paper, we discuss the technoeconomic supplement of P2G with HP for HVDC to explore and utilize renewable energy in the future. We try to answer the following questions of: 1) planning roadmaps of renewable energy, including both wind and solar energy, HVDCs, HPs and P2Gs in future decades; 2) operating combinations of centralized electricity transmission via HVDC and distributed P2G; and 3) the technical and economic advantages of P2G technology in future renewable energy systems. On this basis, a coordinated renewable energy, transmission (including both HVDC and HP) and P2G planning model is required.

## 2. Literature Review

Since there is little existing research on coordinated generation, transmission and P2G planning, the literature review is divided into two parts: research on coordinated generation and transmission planning and research on P2G planning.

Extensive research has been performed on coordinated generation and transmission planning. Most research studies consider only single-stage planning and investment at the beginning of the planning horizon [12-15]. Single-stage planning is the most reasonable approach when dealing with short-time horizons where decisions are not going to be revisited. However, for longer time horizons, multistage planning is essential to closely reproduce the reality of the problem. Multistage planning can consider the trends in the investment costs and the scale of the demand in the long time planning horizon [16], [17].

In addition, the consideration of the uncertainty of renewable energy is an important part of planning models, especially in this research, which studies the sensitivity of HVDC and P2G to this technical factor. Compared to stochastic programming and robust optimization, distributionally robust chance-constrained (DRCC) optimization [18-20] is a kind of uncertainty modeling method that is more suitable for this research: 1) only limited statistical parameters are required which can be obtained from the evaluation results; 2) the conservativeness of the chance constraints is adjustable which is suitable for sensitivity analysis. DRCC optimization has been applied in generation expansion planning [18] and network planning [19], and its effectiveness and advantages have been verified.

Compared to research on coordinated generation and transmission planning, in the literature on P2G planning, many opportunities exist for further modeling improvements. Existing research considers P2G as a kind of energy conversion facility in planning level and describes its model only with the energy conversion efficiency [11], [21-24]. On the one hand, in large-scale application in power systems, P2G should be in the form of clusters rather than a single facility, on the other hand, considering the variability of renewable generation, the start-up and shut-down actions of P2G facilities should be considered, which means the actual operation of P2G clusters should be similar to unit commitment problems [18], [25] of traditional generators. However, a gap remains in the existing studies to describe the unit commitment of P2G at the cluster level.

Above all, due to the lack of satisfactory P2G modeling, there is little research on coordinated HVDC and P2G planning; therefore, the supplement of P2G for HVDC, especially from a technical perspective, is less studied. Based on the existing gap, in this paper, we first propose the complete planning and operation constraints of a P2G cluster. On this basis, the multistage coordinated renewable energy, transmission and P2G planning model is established and applied in the Inner Mongolia-Shandong case in China. The main contributions of this paper are threefold:

1) The complete planning and operation constraints of a P2G cluster considering retirement and unit commitment are first proposed which is essential for research on the large-scale application in renewable energy systems. Furthermore, the "equal-split" rule which determines the power distribution among facilities is verified and applied in model simplification at the cluster level.

2) A multistage coordinated planning model of renewable energy, transmission (HVDC and HP), and P2G is then established which considers multiple energy sectors including electricity, transportation and industry at the same time. In particular, typical characteristics of renewable energy systems are fully considered in the model: the variability of renewable energy is modeled with different scenarios, and uncertainty is modeled as a DRCC program.

3) The proposed coordinated planning model is applied in actual Inner Mongolia-Shandong case studies. The technical and economic advantages of P2G as well as its supplement for HVDC are verified with comparative cases. Furthermore, sensitivity analysis of technical factors (variability and uncertainty) and economic factors (prices and demand) further verifies the importance and limitation of P2G.

The remainder of the paper is organized as follows: Section II describes the complete planning and operation constraints of the P2G cluster. Section III formulates the overall multistage coordinated planning model of renewable energy, transmission and P2G. In Section IV, case studies based on an industrial system of Inner Mongolia-Shandong Province are presented. The summary and conclusions follow in Section V.

## II. Modeling of P2G Cluster

In this section, the complete P2G planning and operation constraints at the cluster level, considering the retirement and unit commitment operation, is established.

### A. Illustration of P2G Cluster

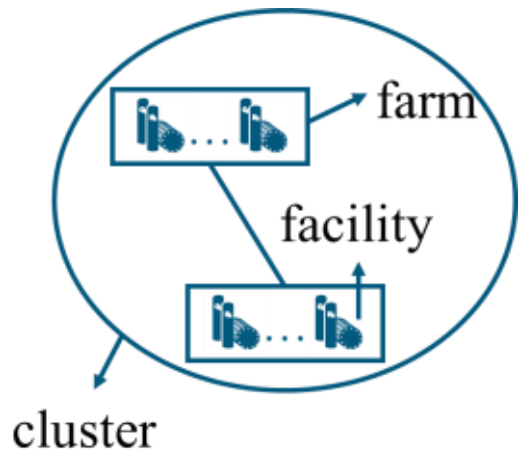

Fig. 1 Illustration of P2G cluster

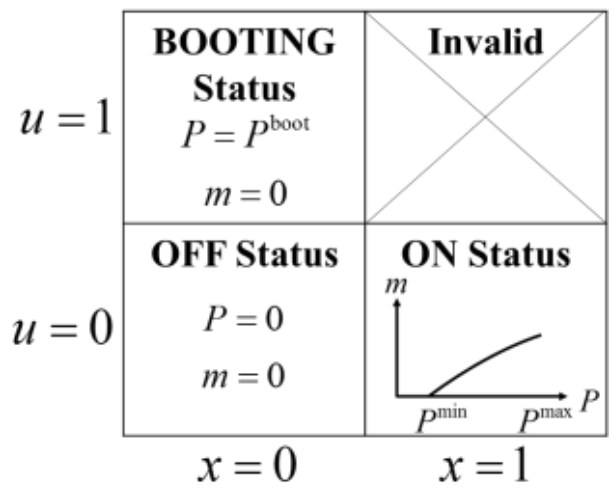

Fig. 2 Three statuses of P2G facilities

Fig. 1 shows the illustration of a P2G cluster. A single P2G facility can attain a maximum of 10 MW (alkaline) at the current stage. For future large-scale applications, we assume a P2G farm consisting of a fixed number of facilities (e.g., 100) as the minimum planning unit. On this basis, a P2G cluster with a larger capacity is formed with several farms. However, the minimum operational unit is still at the facility level. First, the operation status of P2G facilities is introduced, and then the planning and operation constraints at the cluster level is established.

### B. Three Statuses of P2G Facilities

Most existing studies do not consider the different working statuses of the P2G. Considering the variability of renewable energy, P2G can be switched off to save stand-by power $P^{\text{M,min}}$ when there is not enough renewable energy available and switched back on later if needed. According to our previous research [26], there are three statuses in total, as shown in Fig. 2.

1) ON status

When a P2G facility is in ON status, it is an energy conversion unit with adjustable power input and hydrogen output. Based on our previous work [5], the operational flexibility is from the adjustable current $I$ and the temperature $T$. Different working points ($I$, $T$) correspond to different power inputs and hydrogen outputs. Furthermore, due to the overall thermal capacity of P2G, a constraint on the ramping rate exists between two periods. However, when we focus on the operation of the P2G facility from the perspective of power systems, we do not care about detailed operational parameters such as $I$ and $T$; otherwise, we describe $P^{\text{M,min}}$, $P^{\text{M,max}}$, and $\Delta P^{\text{M,max}}$, which can be calculated from operational parameters. Therefore, in ON status, $P_{i,y,s,t}^{\text{M}}$ belongs to [$P^{\text{M,min}}$, $P^{\text{M,max}}$], and the relationship of $m_{i,y,s,t}^{\text{M}}$ with $P_{i,y,s,t}^{\text{M}}$ is a concave function, which can be obtained from the experiment [27]. Here $y = 1,2,...,Y, s = 1,2,...,S, t = 1,2,...,T$. For simpler expressions, we omit the indices explanation in the following equations in this paper.

2) BOOTING status

The shutdown of P2G can be very rapid from the perspective of hydrogen production, and the P2G will be stopped instantaneously once the DC circuit is opened. However, the startup must take some time. No hydrogen can be produced before the stack is heated to the acceptable temperature. Therefore, there is a BOOTING status before the P2G is completely booted, and during this process, a constant power $P^{\text{M,boot}}$ is required for auxiliary facilities with no hydrogen production.

### C. Planning and Operation Constraints of a P2G Cluster

*a. Farm-based Planning Constraints of P2G Cluster*

At the planning level, we consider a P2G farm as the minimum planning unit. Therefore, the planning variable $\delta_{i,y,k}^{\text{M}}$ is the binary variable. We assume that there is an upper limit on the number of new P2G farms $K_{i,y}^{\text{M}}$, and (1) ensures the uniqueness of the planning results. (2) describes the relationship of online P2G farms $\chi_{i,y}^{\text{M}}$ to new P2G farms $\delta_{i,y,k}^{\text{M}}$ considering the lifetime $Y^{\text{M}}$. Considering land eligibility, there is an upper limit on the total number of online P2G farms $K_{i,y}^{\text{M,max}}$ as (3):

$$\delta_{i,y,k+1}^{\text{M}} \le \delta_{i,y,k}^{\text{M}}, 1 \le k < K_{i,y}^{\text{M}} \tag{1}$$

$$\chi_{i,y}^{\text{M}} = \sum_{\tau=y-Y^{\text{M}}+1}^{y} \sum_{k=1}^{K_{i,\tau}^{\text{M}}} \delta_{i,\tau,k}^{\text{M}} \tag{2}$$

$$\chi_{i,y}^{\text{M}} \le K_{i,y}^{\text{M,max}} \tag{3}$$

*b. Facility-based Operation Constraints of P2G Cluster*

Although the planning of the P2G cluster is based on farms, at the operation level, the operation status of each facility is isolated. A feasible working power of the P2G cluster is corresponding to the working status and power of every P2G facility, which reveals the unit commitment of facilities in a cluster should be considered. However, if state variables $x^{\mathrm{M}}_{i,y,s,t}$ and $u^{\mathrm{M}}_{i,y,s,t}$ of each P2G facility are included in the model, it would lead to computational problems.

According to the concave characteristics of ON status shown in Fig. 2, for a P2G cluster with the unified P2G facilities type, **the "equal-split" rule** is verified, which means that the power of each P2G facility in ON status should be equally split by the sum power of the P2G cluster. The detailed proof of the "equal-split" rule is illustrated in the Appendix. Based on the "equal-split" rule, at the cluster level, $x^{\mathrm{M}}_{i,y,s,t}$ and $u^{\mathrm{M}}_{i,y,s,t}$ represent the sum number of facilities that are in ON status and BOOTING status, respectively; in this way, only one group of operation variables is required in the operation model, and the sum power of the P2G cluster can be described as (4). Therefore, the overall operation constraints of the P2G cluster are as follows:

$$\begin{gathered} P^{\mathrm{M}}_{i,y,s,t} = x^{\mathrm{M}}_{i,y,s,t} P^{\mathrm{M,ON}}_{i,y,s,t} + u^{\mathrm{M}}_{i,y,s,t} P^{\mathrm{M,boot}} \\ P^{\mathrm{M,min}} \le P^{\mathrm{M,ON}}_{i,y,s,t} \le P^{\mathrm{M,max}} \end{gathered} \tag{4}$$

$$P^{\mathrm{M}}_{i,y,s,t+1} - P^{\mathrm{M}}_{i,y,s,t} \le x^{\mathrm{M}}_{i,y,s,t} \Delta P^{\mathrm{M,max}} + (\chi^{\mathrm{M}}_{i,y} N^{\mathrm{M}} - x^{\mathrm{M}}_{i,y,s,t}) P^{\mathrm{M,max}} \tag{5}$$

$$m^{\mathrm{M}}_{i,y,s,t} = x^{\mathrm{M}}_{i,y,s,t} \left( a^{\mathrm{M}} (P^{\mathrm{M,ON}}_{i,y,s,t})^2 + b^{\mathrm{M}} (P^{\mathrm{M,ON}}_{i,y,s,t}) + c^{\mathrm{M}} \right) \tag{6}$$

$$x^{\mathrm{M}}_{i,y,s,t} + u^{\mathrm{M}}_{i,y,s,t} \le \chi^{\mathrm{M}}_{i,y} N^{\mathrm{M}} \tag{7}$$

where (4) describes the sum power of the P2G cluster $P^{\mathrm{M}}_{i,y,s,t}$, (5) describes the ramping constraint of P2G cluster, (6) calculates the sum hydrogen production $m^{\mathrm{M}}_{i,y,s,t}$ where the concave function in Fig. 2 can be fitted with the quadratic function $a^{\mathrm{M}}$, $b^{\mathrm{M}}$ and $c^{\mathrm{M}}$, (7) means that the total number of P2G facilities in ON status and BOOTING status should not exceed the total number of online P2G facilities.

Above all, (1)-(7) consist of the complete planning and operation constraints of the P2G cluster. In the following sections, P2G is the abbreviation of P2G cluster without special instructions.

## III. Multistage Coordinated Planning Model of Renewable Energy, Transmission and P2G

This section introduces the multistage coordinated planning model of renewable energy, transmission and P2G. First, the description and assumption of the whole model are illustrated, and then planning constraints on renewable energy, transmission and P2G are explained. The uncertainty of renewable energy output as well as overall operation constraints are modeled with the DRCC. Finally, the objective function is defined.

### A. Model Description and Assumption

As described in the Introduction, we illustrate the problem and the configuration of the coordinated planning model, as shown in Fig. 3. In following expressions, *i* represents source region and *j* represents demand region.

The coordinated planning model is based on the following assumptions [28], [29]:

1) Energy transmission loss via HVDC and energy consumption by compressors of HP are considered in the form of operation cost.

2) The recovery of the residual value of facilities is not considered. During the whole planning horizon, only the replacement of P2G farms is considered since the lifetimes of REs, HVDCs and HPs are generally no less than 30 years.

3) All the facilities are constructed and put into production in the first year in each planning epoch.

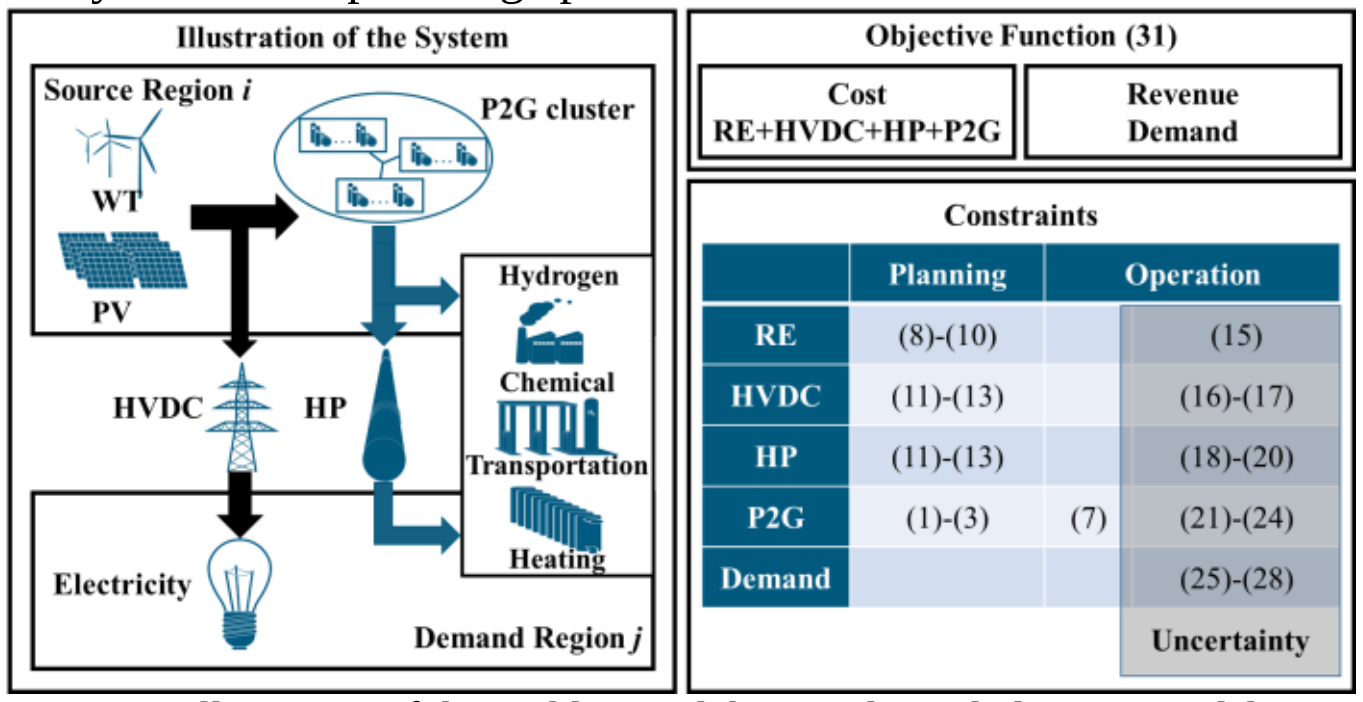

Fig. 3 Illustration of the problem and the coordinated planning model

### B. Coordinated planning model

*a. Planning Constraints of Renewable Energy*

In source region *i*, there are upper limits on the new planning capacity of wind turbines (8), photovoltaics (9), and their sum (10) in each planning epoch *y* and the whole planning horizon:

$$0 \le P^{\mathrm{WT}}_{i,y} \le P^{\mathrm{WT,max}}_{i,y}, \sum_{y=1}^{Y} P^{\mathrm{WT}}_{i,y} \le P^{\mathrm{WT,max}}_{i} \tag{8}$$

$$0 \le P^{\mathrm{PV}}_{i,y} \le P^{\mathrm{PV,max}}_{i,y}, \sum_{y=1}^{Y} P^{\mathrm{PV}}_{i,y} \le P^{\mathrm{PV,max}}_{i} \tag{9}$$

$$0 \le P^{\mathrm{WT}}_{i,y} + P^{\mathrm{PV}}_{i,y} \le P^{\mathrm{max}}_{i,y}, \sum_{y=1}^{Y} (P^{\mathrm{WT}}_{i,y} + P^{\mathrm{PV}}_{i,y}) \le P^{\mathrm{max}}_{i} \tag{10}$$

*b. Planning Constraints of Transmissions*

Here, we consider two kinds of transmissions: HVDC transmission lines and hydrogen transmission pipelines. In each epoch, there is an upper limit on the new planning number $L_{ij,y}$, and there is also an upper limit on the total planning number $L^{\mathrm{max}}_{ij}$ in the planning horizon (11). (12) ensures the uniqueness of the planning results.

$$\sum_{y=1}^{Y} \sum_{l=1}^{L_{ij,y}} \sigma_{ij,y,l} \le L^{\mathrm{max}}_{ij} \tag{11}$$

$$\sigma_{ij,y,l+1} \le \sigma_{ij,y,l}, 1 \le l < L_{ij,y} \tag{12}$$

$$\begin{gathered} \sigma_{ij,y,l} \in \{0,1\}, 1 \le l \le L_{ij,y} \\ ij \in \Omega^{\mathrm{HVDC}} \cup \Omega^{\mathrm{HP}}, y = 1,2,...,Y \end{gathered} \tag{13}$$

*c. Uncertainty Description*

As previously introduced, a DRCC model is applied to address uncertainties from renewable energy. In one source region, there are two kinds of uncertainty sources (wind and solar). The power generation of uncertainty sources is modeled by $\boldsymbol{p}_{s,t} + \boldsymbol{\gamma}_{s,t}$, where $\boldsymbol{p}_{s,t} \in \mathbb{R}^2$ is the mean forecast of wind and solar uncertainty which is a 2-dimensional vector. The forecast error $\boldsymbol{\gamma}_{s,t} \in \mathbb{R}^2$ under a given probability distribution is a random variable with the mean vector $\boldsymbol{\mu}_{s,t} \in \mathbb{R}^2$ and the covariance matrix $\boldsymbol{\Sigma}_{s,t} \in \mathbb{R}^{2\times 2}$, which describes the spatial correlation of different uncertainty sources. It is assumed that exact values for the first- and second-order moments can be estimated from evaluation data [18], and like [18], here, we assume that the mean vector of forecast error is zero, i.e., $\boldsymbol{\mu}_{s,t}=\mathbf{0},\forall s,t$ .

The ambiguity set $\mathbb{P}_{s,t}$ associated with scenario *s* and period *t* is written as:

$$\mathbb{P}_{s,t}=\{D_{s,t}\in\Psi_{s,t}(\mathbb{R}^2):\mathbb{E}^D(\boldsymbol{\gamma}_{s,t})=\boldsymbol{\mu}_{s,t},\mathbb{E}^D(\boldsymbol{\gamma}_{s,t}\boldsymbol{\gamma}_{s,t}{}^{\mathrm{T}})=\boldsymbol{\Sigma}_{s,t}\} \quad (14)$$

where $D_{s,t}$ is a probability distribution belonging to the family of distributions $\Psi_{s,t}(\mathbb{R}^2)$, all with the same first- and second-order moments, including the uncertainty information for all uncertainty sources. Note that $\mathbb{E}^D$ is the expectation operator wherein the uncertainty parameter $\boldsymbol{\gamma}_{s,t}$ follows distribution $D_{s,t}$ . For notational convenience, the indices of $D_{s,t}$ are dropped in the rest of the paper. The complete operation model in the form of DRCC is as follows:

*d. Operation Constraints of Renewable Energy with DRCC*

$$\begin{aligned}P^{\mathrm{WT,E}}_{i,y,s,t}(\boldsymbol{\gamma}_{s,t})+P^{\mathrm{WT,H}}_{i,y,s,t}(\boldsymbol{\gamma}_{s,t})&=\sum_{\tau=1}^{y}P^{\mathrm{WT}}_{i,\tau}\,(p^{\mathrm{WT}}_{i,s,t}(\boldsymbol{\gamma}_{s,t}))\\ P^{\mathrm{PV,E}}_{i,y,s,t}(\boldsymbol{\gamma}_{s,t})+P^{\mathrm{PV,H}}_{i,y,s,t}(\boldsymbol{\gamma}_{s,t})&=\sum_{\tau=1}^{y}P^{\mathrm{PV}}_{i,\tau}\,(p^{\mathrm{PV}}_{i,s,t}(\boldsymbol{\gamma}_{s,t}))\end{aligned} \quad (15)$$

(15) shows that in source region *i*, the renewable generation of wind/solar power can be divided into two parts: electricity transmission via HVDC and hydrogen production via P2G.

*e. Operation Constraints of Transmission with DRCC*

For HVDC transmission lines, power flow is the sum of power for electricity transmission from both wind and solar as in (16).

$$f^{\mathrm{HVDC}}_{ij,y,s,t}(\boldsymbol{\gamma}_{s,t})=P^{\mathrm{WT,E}}_{i,y,s,t}(\boldsymbol{\gamma}_{s,t})+P^{\mathrm{PV,E}}_{i,y,s,t}(\boldsymbol{\gamma}_{s,t}) \quad (16)$$

In this work, the HVDC transmission line is modeled as a link that carries active power within its power limits as a function of possible investments [29]. Hence, the constraint to limit the power flow in HVDC corridors in terms of the investment variables $\sigma_{ij,y,l}$ is represented in (17).

$$\min_{D\in\mathbb{P}_{s,t}}\Pr[\,f^{\mathrm{HVDC}}_{ij,y,s,t}(\boldsymbol{\gamma}_{s,t})\le\sum_{\tau=1}^{y}\sum_{l=1}^{L^{\mathrm{HVDC}}_{ij,y}}\sigma_{ij,y,l}\,f^{\mathrm{HVDC,max}}_{ij}]\ge1-\varepsilon,\ \ ij\in\Omega^{\mathrm{HVDC}} \quad (17)$$

Similarly, the hydrogen input rate of pipelines in source side should not exceed the online maximum flow rate which is determined by compressors, as shown in (18):

$$\min_{D\in\mathbb{P}_{s,t}}\Pr[\sum_{\mathrm{U}}m^{\mathrm{D,U}}_{i,y,s,t}(\boldsymbol{\gamma}_{s,t})\le\sum_{\tau=1}^{y}\sum_{l=1}^{L^{\mathrm{HP}}_{ij,y}}\sigma_{ij,y,l}m^{\mathrm{HP}}_{ij}]\ge1-\varepsilon,\ \ ij\in\Omega^{\mathrm{HP}} \quad (18)$$

Considering the buffer from line packing [30], the operation constraint of the pipeline is shown in (19).

$$m^{\mathrm{HP}}_{ij,y,s,t+1}(\boldsymbol{\gamma}_{s,t+1})=m^{\mathrm{HP}}_{ij,y,s,t}(\boldsymbol{\gamma}_{s,t})+\sum_{\mathrm{U}}m^{\mathrm{D,U}}_{i,y,s,t}(\boldsymbol{\gamma}_{s,t})-\sum_{\mathrm{U}}m^{\mathrm{D,U}}_{j,y,s,t}(\boldsymbol{\gamma}_{s,t}),\ \ ij\in\Omega^{\mathrm{HP}} \quad (19)$$

And $m^{\mathrm{HP}}_{ij,y,s,t}$ should not exceed the maximum storage capacity of the pipeline, as shown in (20).

$$\min_{D\in\mathbb{P}_{s,t}}\Pr[m^{\mathrm{HP}}_{ij,y,s,t}(\boldsymbol{\gamma}_{s,t})\le\sum_{\tau=1}^{y}\sum_{l=1}^{L^{\mathrm{HP}}_{ij,y}}\sigma_{ij,y,l}m^{\mathrm{HP,max}}_{ij}]\ge1-\varepsilon,\ \ ij\in\Omega^{\mathrm{HP}} \quad (20)$$

*f. Operation Constraints of P2G with DRCC*

The P2G power is the sum of the power for hydrogen production from both wind and solar energy.

$$P^{\mathrm{WT,H}}_{i,y,s,t}(\boldsymbol{\gamma}_{s,t})+P^{\mathrm{PV,H}}_{i,y,s,t}(\boldsymbol{\gamma}_{s,t})=\sum_{\mathrm{M}}P^{\mathrm{M}}_{i,y,s,t}(\boldsymbol{\gamma}_{s,t}) \quad (21)$$

Note that the status variables of P2G $x^{\mathrm{M}}_{i,y,s,t}$ and $u^{\mathrm{M}}_{i,y,s,t}$ are independent of short-term uncertainty, i.e., these decisions are made before the time that the uncertainty is realized. Therefore, the operation constraint (7) on these variables is the same as the equations in Section II. However, other operation variables are required to respond to uncertainty, and the operation model of P2G with DRCC is shown below:

$$P^{\mathrm{M}}_{i,y,s,t}(\boldsymbol{\gamma}_{s,t})=x^{\mathrm{M}}_{i,y,s,t}P^{\mathrm{M,ON}}_{i,y,s,t}(\boldsymbol{\gamma}_{s,t})+u^{\mathrm{M}}_{i,y,s,t}P^{\mathrm{M,boot}} \quad (22)$$

$$\min_{D\in\mathbb{P}_{s,t}}\Pr[P^{\mathrm{M}}_{i,y,s,t+1}(\boldsymbol{\gamma}_{s,t+1})-P^{\mathrm{M}}_{i,y,s,t}(\boldsymbol{\gamma}_{s,t})\le x^{\mathrm{M}}_{i,y,s,t}\Delta P^{\mathrm{M,max}}+(\chi^{\mathrm{M}}_{i,y}N^{\mathrm{M}}-x^{\mathrm{M}}_{i,y,s,t})P^{\mathrm{M,max}}]\ge1-\varepsilon \quad (23)$$

$$m^{\mathrm{M}}_{i,y,s,t}(\boldsymbol{\gamma}_{s,t})=x^{\mathrm{M}}_{i,y,s,t}\left(b^{\mathrm{M}}(P^{\mathrm{M,ON}}_{i,y,s,t}(\boldsymbol{\gamma}_{s,t}))+c^{\mathrm{M}}\right) \quad (24)$$

where (22) corresponds to the power constraint (4), (23) corresponds to the ramping constraint (5), and (24) corresponds to the energy conversion constraint (6) , since the concavity is not strong ( $a^{\mathrm{M}}$ is small), for further model simplification, the linear relationship is considered.

*g. Operation Constraints of Demand with DRCC*

For electricity demand on the demand side, it can be described by typical load profiles which is supplied by HVDC as in (25):

$$f^{\mathrm{HVDC}}_{ij,y,s,t}(\boldsymbol{\gamma}_{s,t})=P^{\mathrm{D}}_{j,y,s}(\boldsymbol{\gamma}_{s,t})\,p^{\mathrm{D}}_{j,s,t} \quad (25)$$

For hydrogen demand on both source side and demand side, considering the differences of hydrogen requirements in different hydrogen sectors, annual upper limits in each sector are required:

$$\sum_{\mathrm{U}}(m^{\mathrm{S,U}}_{i,y,s,t}(\boldsymbol{\gamma}_{s,t})+m^{\mathrm{D,U}}_{i,y,s,t}(\boldsymbol{\gamma}_{s,t}))=\sum_{\mathrm{M}}m^{\mathrm{M}}_{i,y,s,t}(\boldsymbol{\gamma}_{s,t}) \quad (26)$$

$$\min_{D\in\mathbb{P}_{s,t}}\Pr[\sum_{s=1}^{S}D_s\sum_{t=1}^{T}(m^{\mathrm{S,U}}_{i,y,s,t}(\boldsymbol{\gamma}_{s,t}))\le m^{\mathrm{S,U}}_{i,y}]\ge1-\varepsilon \quad (27)$$

$$\min_{D\in\mathbb{P}_{s,t}}\Pr[\sum_{s=1}^{S}D_s\sum_{t=1}^{T}(m^{\mathrm{D,U}}_{j,y,s,t}(\boldsymbol{\gamma}_{s,t}))\le m^{\mathrm{D,U}}_{j,y}]\ge1-\varepsilon \quad (28)$$

where (26) describes the hydrogen balance between supply and demand, and U represents hydrogen downstream sectors. Here, we consider three main sectors: chemical (C), transportation (T) and heating (H). According to the prediction of future requirements, there are upper limits on requirements in each sector in source (S) and demand (D) regions, which are described by (27)-(28).

*h. Model Simplification*

To further mitigate the complexity of the proposed model, the continuous operation variables are simplified with linear decision rules [18]. In this way, all the continuous operation variables $cov_{s,t}$ in (15)-(28) can be described as follows:

$$cov_{s,t}(\boldsymbol{\gamma}_{s,t})=cov_{s,t}+\beta cov_{s,t}(\mathbf{1}^{\mathrm{T}}\boldsymbol{\gamma}_{s,t}) \quad (29)$$

Note that $cov_{s,t}$ is the tentative schedule value based on the prediction, whereas $\beta cov_{s,t}(\mathbf{1}^{\mathrm{T}}\boldsymbol{\gamma}_{s,t})$ represents the linear response to the uncertainty renewable power generation.

By adopting Cantelli's inequality (a one-sided Chebyshev inequality), the distributionally robust chance inequality constraints can be formulated as second-order cone constraints.

$$\min_{D\in\mathbb{P}_{s,t}}\Pr[\boldsymbol{A}_{s,t}{}^{\mathrm{T}}\boldsymbol{\gamma}_{s,t}\le b_{s,t}]\ge1-\varepsilon\Leftrightarrow\sqrt{\boldsymbol{A}_{s,t}{}^{\mathrm{T}}\boldsymbol{\Sigma}_{s,t}\boldsymbol{A}_{s,t}}\le\sqrt{\frac{\varepsilon}{1-\varepsilon}}b_{s,t} \quad (30)$$

With linear decision rules and Cantelli's inequality, the complete operation model can be reformulated into a MISOCP form, and the detailed equations are in the Appendix.

*i. Objective Function*

From the prospective of the government who concerns how to utilize renewable energy to help the decarbonization of energy sectors, the overall economy of coordinated renewable energy, transmission and P2G planning should be analyzed. Therefore, the objective function (31) includes two main parts: cost of renewable energy generators $C^{\mathrm{RE}}$, HVDCs $C^{\mathrm{HVDC}}$, HPs $C^{\mathrm{HP}}$, P2Gs $C^{\mathrm{P2G}}$, and revenue from both electricity $R^{\mathrm{E}}$ and hydrogen $R^{\mathrm{H}}$. Both cost and revenue are net present value:

$$J = \min C^{\mathrm{RE}} + C^{\mathrm{HVDC}} + C^{\mathrm{HP}} + C^{\mathrm{P2G}} - R^{\mathrm{E}} - R^{\mathrm{H}} \tag{31}$$

For renewable energy generators, HVDCs, HPs and P2Gs, we consider mainly the capital cost and fixed operation cost, and booting cost for P2Gs is the variable operation cost. They are all independent of short-term uncertainty from renewable energy.

For revenue from electricity and hydrogen, $R^{\mathrm{E}}$ is related to $P^{\mathrm{D}}_{j,y,s}(\gamma_{s,t})$, and $R^{\mathrm{H}}$ is from $m^{\mathrm{S,U}}_{i,y,s,t}$ and $m^{\mathrm{D,U}}_{j,y,s,t}$. Detailed expressions of (31) are in the Appendix.

## IV. Case Studies

In this section, case studies are performed based on the actual industrial system of Inner Mongolia-Shandong Province. First proposed coordinated planning model is applied, and the supplement of P2G for HVDC is verified from the perspective of both technology and economy. Furthermore, sensitivity analysis on technical factors (variability and uncertainty) and economic factors (prices and demand) further verifies the advantages and limitations of P2G.

The proposed coordinated planning model is a MISOCP optimization problem that is coded in MATLAB and solved with CPLEX12.6.

### *A. Case Descriptions*

The proposed model is applied in the typical case of Inner Mongolia and Shandong Province in China. According to renewable energy evaluation based on GREAN platform [31], there are wind and solar resources in the Bayannaoer region in Inner Mongolia. At the same time, there is distributed hydrogen demand in Inner Mongolia from the chemical, transportation and heating sectors. For Shandong Province, which is approximately 1230 km from Bayannaoer, both electricity demand and hydrogen demand can be satisfied via HVDC and HP by renewable energy in Bayannaoer.

Simple parameters in the model are shown directly in the nomenclature, and parameter-related P2Gs are obtained from [32] and [33]. In this case study, we discuss only alkaline technology. Other economic parameters in (31) and hydrogen demand parameters in (27)-(28) according to the research on realistic data of Inner Mongolia and Shandong are shown in Table I and Fig. 4. We assume that the prices of facilities decrease linearly in planning horizon.

Table I
Parameters in Coordinated Planning Model

| Epoch | $c^{\mathrm{WT}}_y$ | $c^{\mathrm{PV}}_y$ | $c^{\mathrm{M}}_y$ | $c^{\mathrm{E}}_y$ | $c^{\mathrm{S,U}}_y$ / $c^{\mathrm{D,U}}_y$ (RMB/kg) | | |
|---|---|---|---|---|---|---|---|
| | (RMB/kW) | | | (RMB/kWh) | C | T | H |
| 2020 ($y$=1) | 7500 | 4200 | 2400 | 0.4 | 12/15.5 | 30 | 17.5 |
| 2050 ($y$=6) | 3600 | 1600 | 1200 | 0.2 | | | |

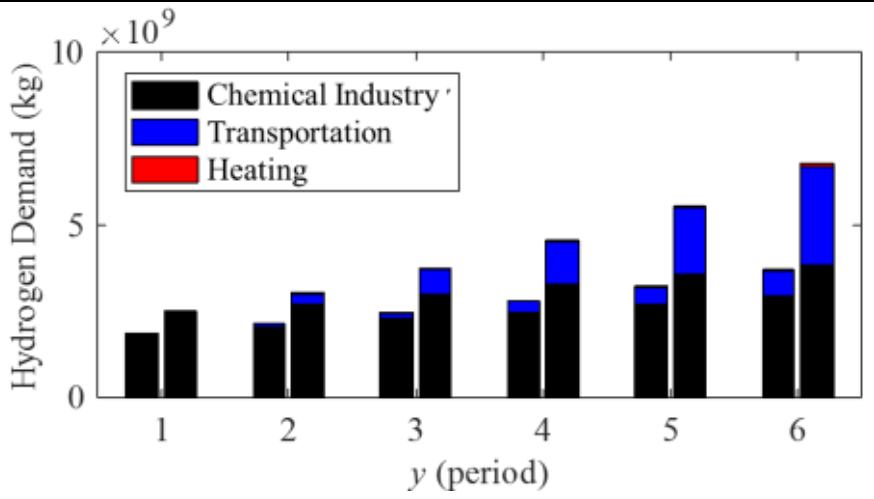

Fig. 4 Hydrogen demand in Inner Mongolia(left)/Shandong(right)

The profiles of wind and solar based on the evaluation and load are in the Appendix. Four scenarios are considered in the benchmark case. These scenarios are generated by K-means clustering of 365 scenarios, and $D_s$ of each scenario can be obtained at the same time.

### *B. Supplement of P2G with HP for HVDC*

*a. Economic Supplement*

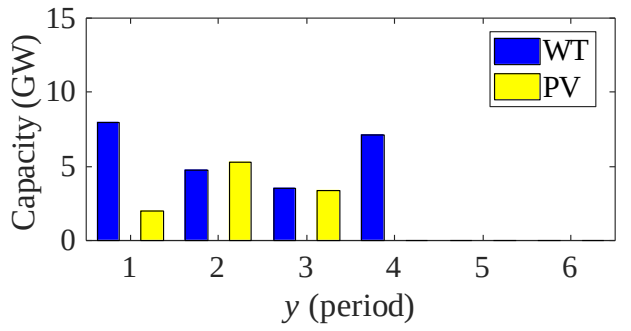

Fig. 5 Multistage planning result of REs

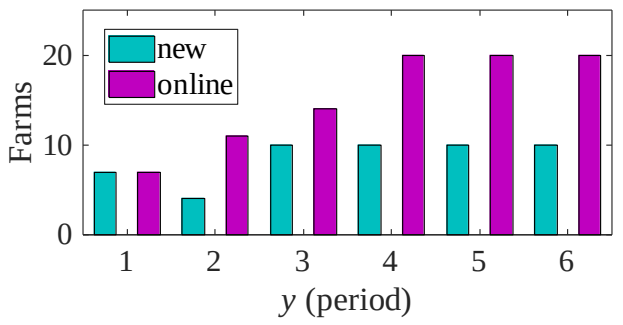

Fig. 6 Multistage planning result of P2Gs

Fig. 5 and Fig. 6 show the multistage planning results of REs and P2Gs, respectively. In addition, HVDC is established in the first planning epoch, while HP is established in the second planning epoch. Finally, in the whole planning horizon, 23 GW WT, 11 GW PV, 8GW HVDC, 48inch HP and maximal 20 GW online P2Gs are established.

In Table II and the following discussion, the letter E represents energy for electricity transmission, and the letter H represents energy for hydrogen production. Table II shows that following the planning orders of HVDC and P2Gs, the ratio of E to H gradually decreases, and in the last epoch, E only accounts for 23% of the total renewable energy. The ratio of E to H also influences the annual utilization hours (AUH) of HVDC, P2G and HP, as shown in Table II.

Table II
Multistage Operation Results

| Epoch | E (TWh) | H (TWh) | $\mathrm{AUH^{RE}}$ | $\mathrm{AUH^{HVDC}}$ | $\mathrm{AUH^{HP}}$ | $\mathrm{AUH^{P2G}}$ |
|---|---|---|---|---|---|---|
| 1 | 22 (65%) | 11 | 3400 | 2804 | 0 | 1653 |
| 2 | 24 (39%) | 37 | 3081 | 3044 | 2174 | 3388 |
| 3 | 24 (29%) | 56 | 3020 | 3044 | 3070 | 4064 |
| 4-6 | 24 (23%) | 83 | 3181 | 3044 | 5345 | 4190 |

In summary, the planning of P2Gs and HPs are later than HVDC, and the ratio of E to H is gradually decreasing, which due mainly to 1) $c^{\mathrm{WT}}_y$, $c^{\mathrm{PV}}_y$ and $c^{\mathrm{E}}_y$ decreasing; 2) $c^{\mathrm{M}}_y$ decreasing; and 3) $m^{\mathrm{L,U}}_{i,y}$ and $m^{\mathrm{R,U}}_{j,y}$ gradually increasing especially in the transportation sector. Planning and operation results reveal the economic supplement of P2G with HP for HVDC in long term, and the advantage of hydrogen production compared to electricity transmission gradually become obvious.

*b. Technical Supplement*

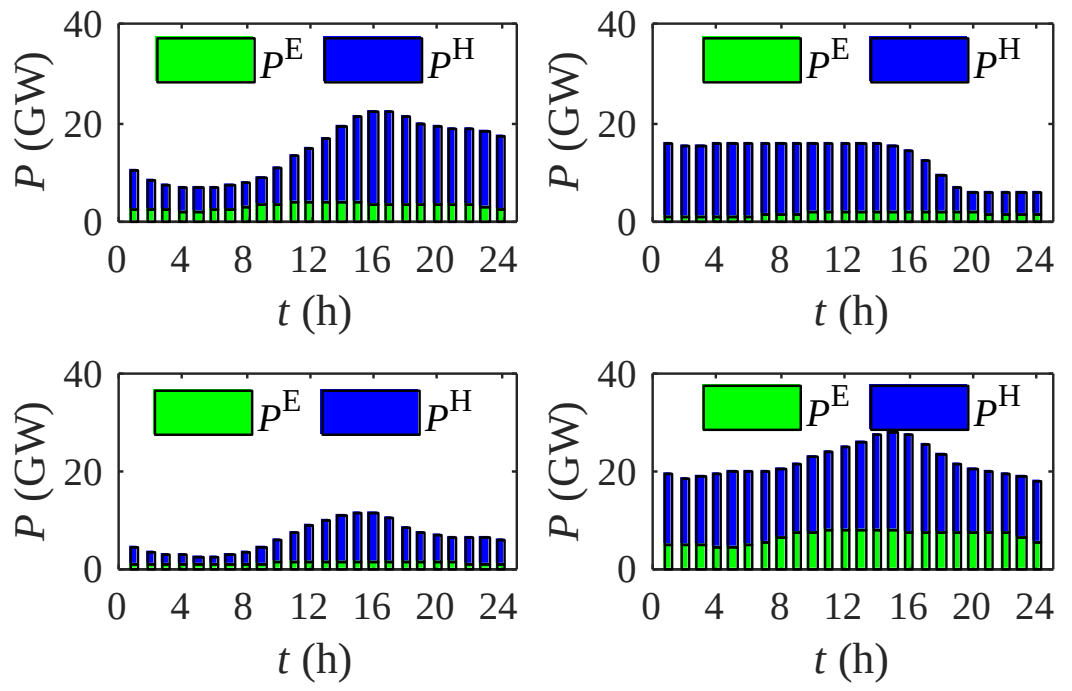

Fig. 7 Optimal operation result of REs in the last epoch

Fig. 7 shows the operation profiles of REs. For renewable energy, there are significant differences between different scenarios, especially for wind. Electricity transmission via HVDC requires the balance of source and demand, therefore P2G with HP provide the necessary operation supplement and buffer for HVDC to consume intraday and interday fluctuations since the transmission and utilization of hydrogen are bufferable. The intraday operation results of P2G cluster is shown in Fig. 8 below.

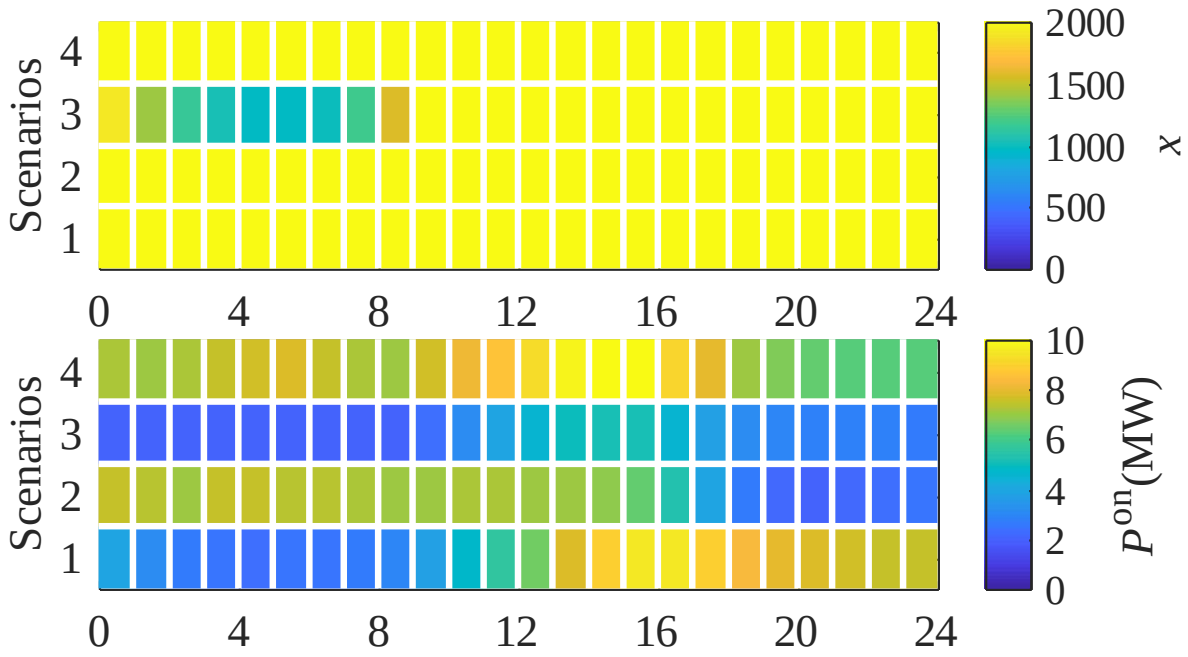

Fig. 8 Optimal operation results of P2G cluster in the last epoch

*c. Economic Analysis*

In the whole planning horizon, the levelized cost of electricity (LCOE), levelized cost of HVDC (LCOHVDC), levelized cost of P2G (LCOP2G), levelized cost of HP (LCOHP), levelized profit of electricity (LPOE) and levelized profit of hydrogen (LPOH) are calculated based on [28]. The results are shown in Fig. 9. P2G(S/D) represents the results for hydrogen demand in the source/demand region, respectively.

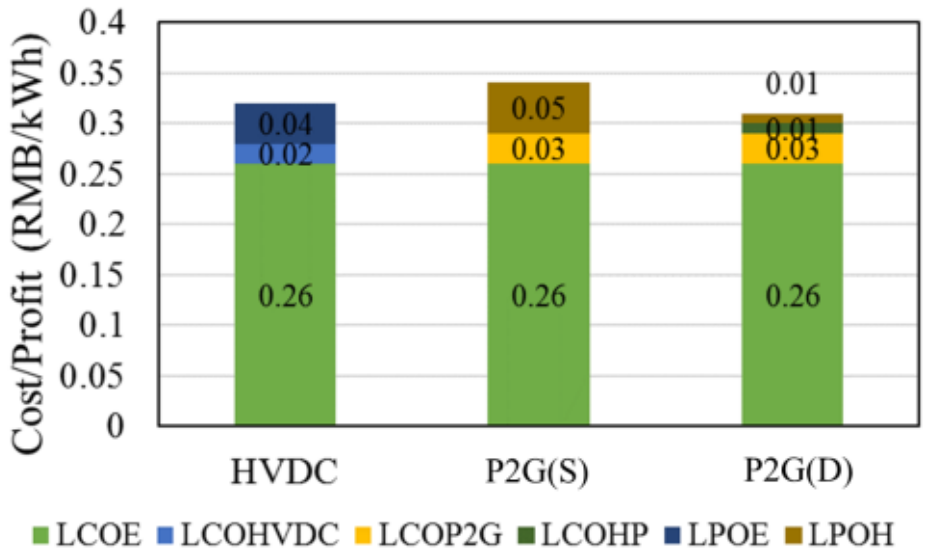

Fig. 9 Economic results of HVDC and P2G

From Fig. 9, the conclusions below are reached:

1) Except for the basic LCOE, the additional levelized cost of the P2G (0.04 RMB/kWh) is approximately **twice** the HVDC (0.02 RMB/kWh).

2) The levelized cost of hydrogen (LCOH) is approximately 0.30 RMB/kWh (20 RMB/kg); therefore, the profit of the P2G depends mainly on hydrogen application in the **transportation sector** (30 RMB/kg).

*d. Comparison with HVDC Only*

Considering the case in a renewable energy system without P2G, and energy can only be consumed via electricity transmission. The planning, operation and economic results are shown in Table III.

TABLE III
COMPARISON OF HVDC+P2G WITH HVDC

| Case | Planning (WT/PV) | Operation | Economy |
|---|---|---|---|
| HVDC | 8 GW/2 GW | Maximum 66% consumption | LCOE=0.33 RMB/kWh Profit=3 billion RMB |
| HVDC +P2G | 23 GW/11 GW | Extra over 34% consumption | LCOE=0.26 RMB/kWh Profit=70 billion RMB |

Compared to the benchmark case, the advantages of P2G can be quantitatively concluded to be:

1) From the perspective of planning, an extra 24 GW of renewable energy can be explored economically.

2) From the perspective of operation, based on the operational flexibility of the P2G cluster, Fig. 8 verifies that P2G is a kind of flexible resource that can cooperate well with the power grid for extra above 34% renewable energy consumption.

3) From the perspective of economy, since the economy of HVDC is becoming worse with the decline of facility investment cost, it demonstrates advantages in the short term, but with the hydrogen required in the transportation sector increasing, P2G in fact occupies advantages in long term, which is an exact complement to HVDC. With more exploration of renewable energy at a lower cost, the present value of LCOE decreases from 0.33 RMB/kWh to 0.26 RMB/kWh, and the total profit in the whole planning horizon increases significantly.

## C. Sensitivity Analysis

In this subsection, sensitivity analysis of technical factors and economic factors is quantitatively studied to verify the advantages and limitations of P2Gs. Considered factors are shown in detail in Table IV. First, the results of the sensitivity analysis are listed in Tables V, VI and VIII, and then factors that are beneficial/unfavorable to P2Gs are identified and discussed.

TABLE IV
FACTORS CONSIDERED IN SENSITIVITY ANALYSIS

| Factors | | Parameters |
|---|---|---|
| Technical | Variability | $S$ |
| | Uncertainty | $\varepsilon$ |
| Economic | Prices of RE | $c_y^{\mathrm{WT}}$ / $c_y^{\mathrm{PV}}$ |
| | Hydrogen demand in transportation sector | $m_{i,y}^{\mathrm{S,T}}$ / $m_{j,y}^{\mathrm{D,T}}$ |

*a. Variability*

TABLE V
SENSITIVITY ANALYSIS RESULTS ON VARIABILITY

| *S* | WT/PV (GW) | HVDC | HP | P2G (new/online) | E (%) (sum) |
|---|---|---|---|---|---|
| 4 | 23/11 | 1 | 1 | 51/20 | 28.93% |
| 6 | 22/12 | 1 | 1 | 51/20 | 28.69% |
| 8 | 19/14 | 1 | 1 | 52/20 | 26.50% |
| 10 | 18/14 | 1 | 1 | 53/20 | 25.96% |

Table V reveals that with the number of scenarios $S$ increasing which represents the stronger variability of renewable energy, following rules can be seen:

1) In planning level, the number of new P2Gs increases, which means that the planning of P2Gs moves up.

2) In operation level, the percentage of energy for electricity transmission (E) decreases mainly due to the technical constraint from the stronger imbalance of source and demand profiles.

Therefore compared to HVDC, P2Gs are more suitable for following renewable energy output with strong variability (such as wind power).

*b. Uncertainty*

TABLE VI
SENSITIVITY ANALYSIS RESULTS ON UNCERTAINTY

| $\varepsilon$ | WT/PV (GW) | HVDC | HP | P2G (GW) (new/online) | E (%) (sum) |
|---|---|---|---|---|---|
| - | 23/11 | 1 | 1 | 51/20 | 29% |
| 0.01 | 12/10 | 1 | 0 | 28/11 | 49% |
| 0.1 | 13/10 | 1 | 0 | 30/11 | 47% |

Table VI shows that with the consideration of the uncertainty of renewable energy, following rules can be seen:

1) In planning level, the planning of HP and P2G are more conservative, HP would be no longer planned.

2) In operation level, the percentage of energy for electricity transmission (E) significantly increases.

It is because compared to HVDC with the large capacity, P2G with small capacity is more sensitive to the uncertainty, besides additional levelized cost of P2G is higher than that of HVDC, therefore in the worst case overinvestment would lead to bad economy.

Furthermore, from Table VI, the smaller $\varepsilon$ is (the more conservative of the DRCC model), the less capacity of P2Gs, and the more energy is transported via HVDC rather than converted into hydrogen, which also reveals the limitation of P2Gs when facing uncertainty.

*c. Economic Factors*

We consider the four comparative cases on economic factors shown in Table VII, here "-" means that in this case parameters are unchanged in the planning horizon and "↓" means that parameters will decline. The results are shown in Table VIII.

TABLE VII
SET OF COMPARATIVE CASES

| Economic factors | Case1 | Case2 | Case3 | Case4 |
|---|---|---|---|---|
| Prices of RE | ↓ | – | ↓ | – |
| Demand in transportation sector | – | – | ↓ | ↓ |

TABLE VIII
SENSITIVITY ANALYSIS RESULTS ON ECONOMIC FACTORS

| Case | WT/PV(RE) (GW) | HVDC | HP | P2G (GW) (new/online) | E (%) (sum) |
|---|---|---|---|---|---|
| 1 | 23.36/10.61(34) | 1 | 1 | 51/20 | 29% |
| 2 | 23.22/10.77(34) | 1 | 1 | 53/20 | 37% |
| 3 | 9.09/3.49(13) | 1 | 0 | 23/8 | 63% |
| 4 | 20.48/4.03(25) | 2 | 0 | 51/18 | 64% |

1) Prices of RE

In planning level, the reduction on cost significantly decreases the economy of HVDC and shrinks the planning of REs, HVDCs and P2Gs especially when the development of hydrogen transportation sector is pessimistic (Case3 and Case4). In operation level, the percentage of energy for electricity transmission (E) decreases with the reduction of cost (Case1 and Case2). The results verify the advantages of P2Gs following the decreasing tendency of RE's cost.

2) Hydrogen demand in transportation sector

Compared Case1, 2 with Case3, 4, if the development of transportation sector is pessimistic, HP is not planned and the number of P2Gs also reduces, and the percentage of energy for electricity transmission (E) significantly increases. It verifies the conclusion from above economic analysis that the profit of P2G mainly depends on hydrogen application in transportation sector.

*d. Advantages and Limitations of P2G*

In summary, both technical factors (variability and uncertainty) and economic factors (prices and demand) influence the planning and operation of HVDCs, HPs and P2Gs, and the correlation can be concluded as shown in Table IX.

TABLE IX
SENSITIVITY ANALYSIS OF TECHNICAL AND ECONOMIC FACTORS

| Factors | | Planning | | Operation |
|---|---|---|---|---|
| | | HVDC | P2G+HP | E/H |
| Technical | Variability ↑ | - | ↑ | ↓ |
| | Uncertainty ↑ | - | ↓ | ↑ |
| Economic | Prices of RE ↓ | ↓ | ↓ | ↓ |
| | Demand of hydrogen ↓ | - | ↓ | ↑ |

From a technical perspective, the weak correlation of HVDC with technical factors shows its strong robustness with large capacity. In contrast, P2G shows its advantages in strong variability with small capacity and operation flexibility (Table V) and its disadvantage in uncertainty (Table VI), due mainly to the higher additional levelized cost, which is approximately twice HVDC. Therefore, in the worst case, the investment would be conservative.

From an economic perspective, the reduction in cost significantly decreases the economy of the HVDC, which is beneficial to hydrogen production (Table VIII). However, when the development of the hydrogen transportation sector is pessimistic, P2G shows limitations in both the planning and operation levels, which verifies the conclusion that the profit of the P2G depends mainly on hydrogen application in the transportation sector.

## V. CONCLUSIONS

Focusing on future large-scale renewable energy utilization, this paper considers two kinds of consumption modes (electricity and hydrogen) and studies the technoeconomic supplement of P2G with HP for HVDC. First the complete planning and operation constraints of a large-capacity P2G cluster considering retirement and unit commitment operation has been proposed. On this basis, a multistage coordinated planning model of REs, HVDCs, HPs, and P2Gs is established considering the variability and uncertainty of renewable energy. The industrial case of Inner Mongolia-Shandong is chosen for case studies.

Multistage planning and operation results show the obvious temporal complementarity in which HVDC and P2G occupy advantages in the short term and long term, respectively. Compared to HVDC alone, P2G can provide both technical and economic supplements: 1) An extra 24 GW of renewable energy can be explored with profit; 2) P2G is a kind of flexible resource that can cooperate well with the power grid for extra (above 34%) renewable energy consumption; 3) With more exploration of renewable energy at a lower cost, the present value of LCOE decreases from 0.33 RMB/kWh to 0.26 RMB/kWh, which gains profits for both HVDV and P2G.

Furthermore, sensitivity analysis on both technical and economic factors further verifies the advantages of P2G: considering the strong variability of renewable energy and downward tendency of facilities' cost, energy is prone to be consumed from HVDC-majored to P2G-majored. However, since the additional levelized cost of the P2G (0.04 RMB/kWh) is approximately twice the HVDC (0.02 RMB/kWh), and the profit of the P2G depends mainly on hydrogen application in the transportation sector, the P2G is more sensitive to the uncertainty from renewable energy and future hydrogen demand.

## APPENDIX

### A. *Proof of "Equal-split" Rule*

For a simpler illustration, $m_i$ and $P_i$ represent the hydrogen production and the power of the P2G facility *i* in ON status in the P2G cluster, and the relationship of $m_i$ and $P_i$ can be described with $f_i$ which is a concave function as (32).

$$m_i = f_i(P_i) \tag{32}$$

$m$ and $P$ represent the sum hydrogen production and the sum power of the P2G cluster, respectively, which can be described as follows:

$$\begin{aligned} &\max m = \sum_i m_i = \sum_i f_i(P_i) \\ &s.t. \sum_i P_i = P \end{aligned} \tag{33}$$

The objective function of the above maximum problem is a concave function, and the equality constraint is linear; therefore, the optimality condition of this problem is:

$$-f_i^{'}(P_i) - \lambda = 0, \forall i, \lambda \in \mathbb{R} \tag{34}$$

where $\lambda$ is the Lagrange multiplier of the equality constraint. Since $\lambda$ is unique, and the optimality condition reveals that for an arbitrary P2G facility in ON status, the optimal $P_i$ should satisfy that all the $f_i^{'}(P_i)$ are the same. Especially when all P2G facilities are in the same type, $P_i$ should be equally split by the sum power of the P2G cluster for maximum hydrogen production, which is the "equal-split" rule.

### B. *Parameters in Models*

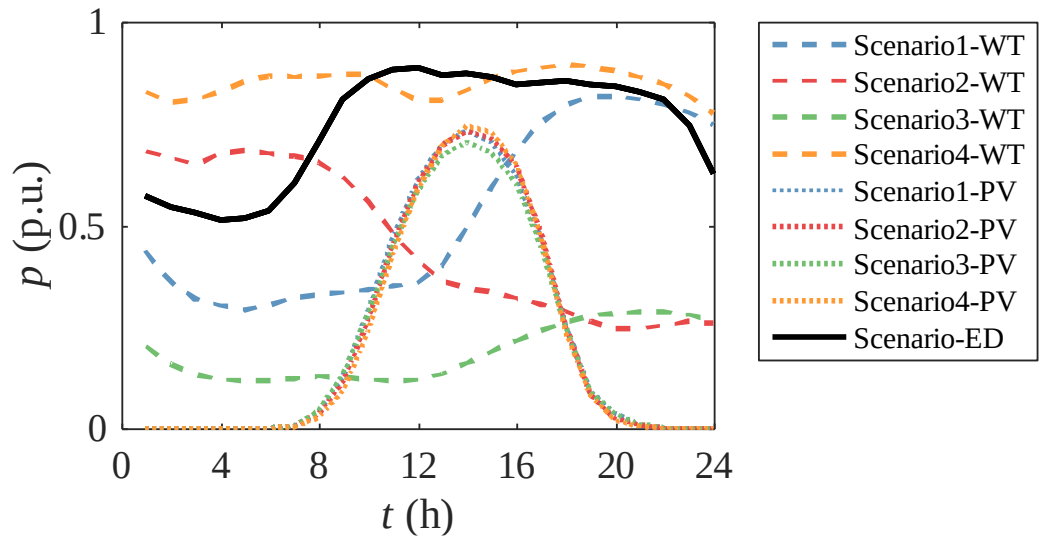

Fig. 10 Typical Profiles of wind turbines, photovoltaics and loads

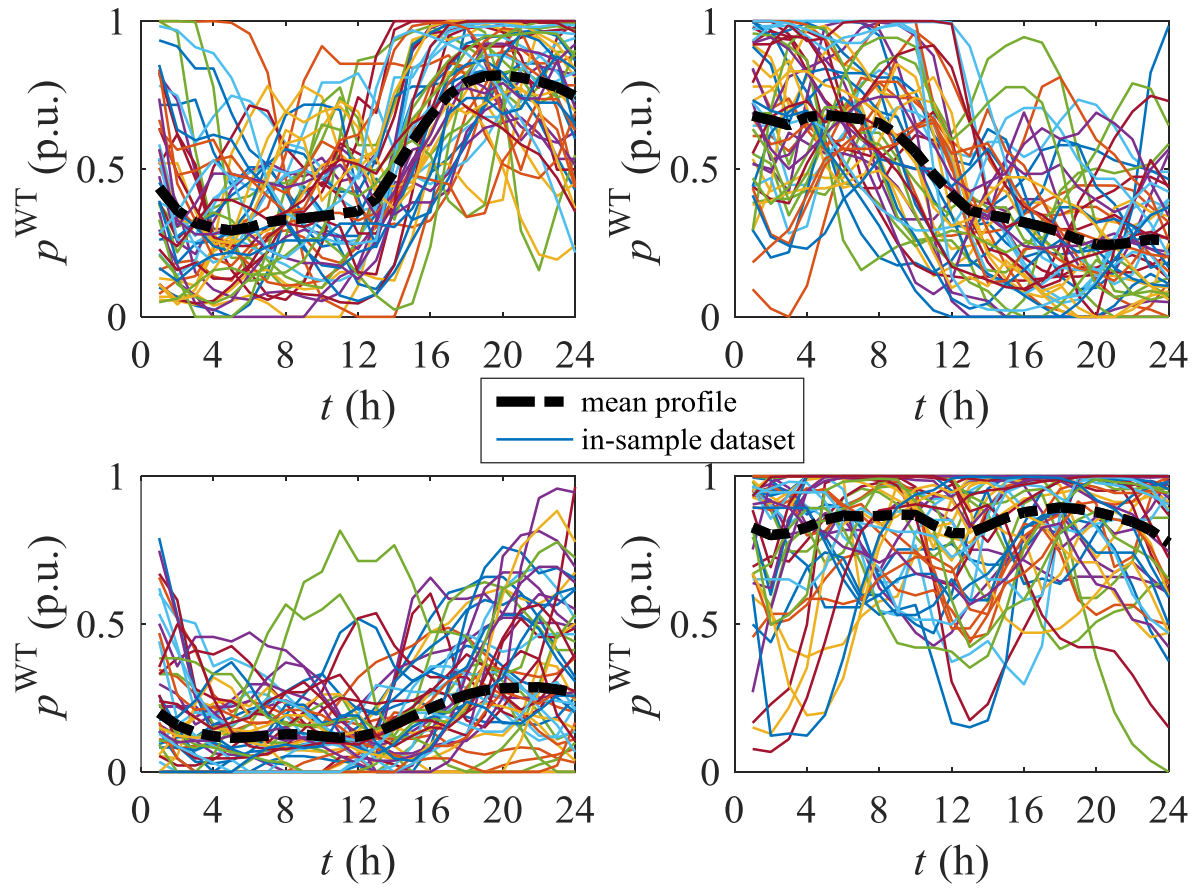

Fig. 11 Typical profiles of the WT and in-sample dataset in the DRCC model

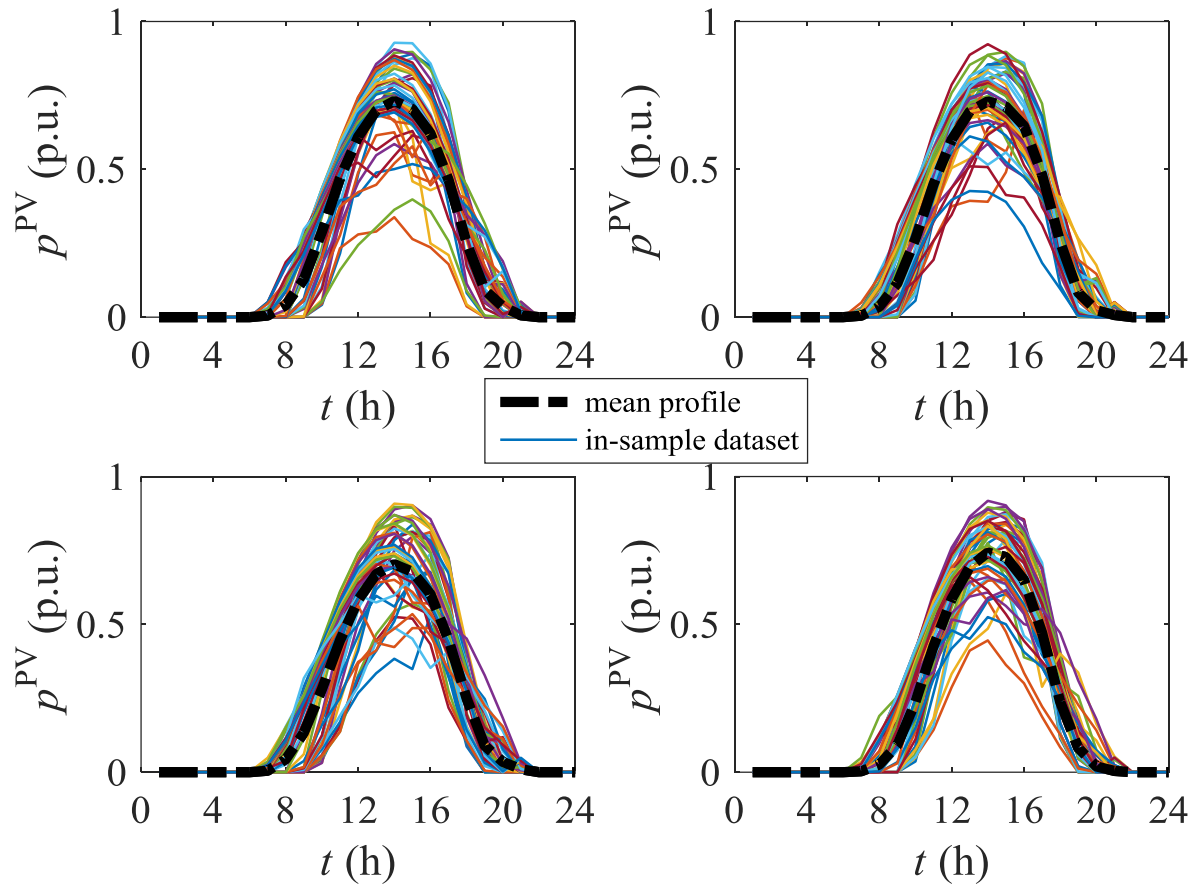

Fig. 12 Typical profiles of the PV and in-sample datasets in the DRCC model

### C. *Detailed Expressions of Objective Functions*

Detailed expressions of (31) are as follows:

$$C^{\mathrm{RE}}=\underbrace{\sum_{y=1}^{Y}\alpha_{(y-1)*epoch+1}(c_y^{\mathrm{WT}}P_{i,y}^{\mathrm{WT}}+c_y^{\mathrm{PV}}P_{i,y}^{\mathrm{PV}})}_{\mathrm{CAPEX^{RE}}}+\underbrace{\sum_{y=1}^{Y}\beta_y(k^{\mathrm{WT}}c_y^{\mathrm{WT}}P_{i,y}^{\mathrm{WT}}+k^{\mathrm{PV}}c_y^{\mathrm{PV}}P_{i,y}^{\mathrm{PV}})}_{\mathrm{OPEX^{RE}}} \quad (35)$$

$$C^{\mathrm{HVDC}}=\underbrace{\sum_{ij\in\Omega^{\mathrm{HVDC}}}\sum_{y=1}^{Y}\alpha_{(y-1)*epoch+1}\left(c_{ij}^{\mathrm{HVDC}}\sum_{l=1}^{L_{ij,y}^{\mathrm{HVDC}}}\sigma_{ij,y,l}\right)}_{\mathrm{CAPEX^{HVDC}}}+\underbrace{\sum_{ij\in\Omega^{\mathrm{HVDC}}}\sum_{y=1}^{Y}\beta_y\left(k^{\mathrm{HVDC}}c_{ij}^{\mathrm{HVDC}}\sum_{l=1}^{L_{ij,y}^{\mathrm{HVDC}}}\sigma_{ij,y,l}\right)}_{\mathrm{OPEX^{HVDC}}} \quad (36)$$

$$C^{\mathrm{HP}}=\underbrace{\sum_{ij\in\Omega^{\mathrm{HP}}}\sum_{y=1}^{Y}\alpha_{(y-1)*epoch+1}\left(c_{ij}^{\mathrm{HP}}\sum_{l=1}^{L_{ij,y}^{\mathrm{HP}}}\sigma_{ij,y,l}\right)}_{\mathrm{CAPEX^{HP}}}+\underbrace{\sum_{ij\in\Omega^{\mathrm{HP}}}\sum_{y=1}^{Y}\beta_y\left(k^{\mathrm{HP}}c_{ij}^{\mathrm{HP}}\sum_{l=1}^{L_{ij,y}^{\mathrm{HP}}}\sigma_{ij,y,l}\right)}_{\mathrm{OPEX^{HP}}} \quad (37)$$

$$\begin{aligned}C^{\mathrm{P2G}}=&\underbrace{\sum_{y=1}^{Y}\alpha_{(y-1)*epoch+1}\sum_{\mathrm{M}}c_y^{\mathrm{M}}\sum_{k=1}^{K_{i,y}^{\mathrm{M}}}\delta_{i,y,k}^{\mathrm{M}}}_{\mathrm{CAPEX^{P2G}}}+\underbrace{\sum_{y=1}^{Y}\beta_y\sum_{\mathrm{M}}k^{\mathrm{M}}c_y^{\mathrm{M}}\sum_{k=1}^{K_{i,y}^{\mathrm{M}}}\delta_{i,y,k}^{\mathrm{M}}}_{\mathrm{fixOPEX^{P2G}}}+\dots\\&\underbrace{\sum_{y=1}^{Y}\beta_y\sum_{s=1}^{S}D_s\sum_{t=1}^{T}\sum_{\mathrm{M}}c^{\mathrm{M,boot}}u_{i,y,s,t}^{\mathrm{M}}}_{\mathrm{varOPEX^{P2G}}}\end{aligned} \quad (38)$$

$$\begin{aligned}R^{\mathrm{E}}&=\sum_{ij\in\Omega^{\mathrm{HVDC}}}\sum_{y=1}^{Y}c_y^{\mathrm{E}}\beta_y\max_{D\in\mathbb{P}_{s,t}}\sum_{s=1}^{S}D_s\sum_{t=1}^{T}\min\mathbb{E}^{D}[P_{j,y,s}^{\mathrm{D}}(\boldsymbol{\gamma}_{s,t})p_{j,s,t}^{\mathrm{D}}]\\&=\sum_{ij\in\Omega^{\mathrm{HVDC}}}\sum_{y=1}^{Y}c_y^{\mathrm{E}}\beta_y\sum_{s=1}^{S}D_s\sum_{t=1}^{T}P_{j,y,s}^{\mathrm{D}}p_{j,s,t}^{\mathrm{D}}\end{aligned} \quad (39)$$

$$\begin{aligned}R^{\mathrm{H}}&=\sum_{y=1}^{Y}\beta_y\max_{D\in\mathbb{P}_{s,t}}\sum_{s=1}^{S}D_s\sum_{t=1}^{T}\min\mathbb{E}^{D}[\sum_{\mathrm{U}}c_y^{\mathrm{S,U}}m_{i,y,s,t}^{\mathrm{S,U}}(\boldsymbol{\gamma}_{s,t})+c_y^{\mathrm{D,U}}m_{i,y,s,t}^{\mathrm{D,U}}(\boldsymbol{\gamma}_{s,t})]\\&=\sum_{y=1}^{Y}\beta_y\sum_{s=1}^{S}D_s\sum_{t=1}^{T}\sum_{\mathrm{U}}c_y^{\mathrm{S,U}}m_{i,y,s,t}^{\mathrm{S,U}}+c_y^{\mathrm{D,U}}m_{i,y,s,t}^{\mathrm{D,U}}\end{aligned} \quad (40)$$

where $\beta_y=\alpha_{(y-1)*epoch+1}+\dots+\alpha_{y*epoch}$, $\alpha_y=\dfrac{1}{(1+r)^y}$ is the present value interest factor.

Considering the uncertainty, $R^{\mathrm{E}}$ and $R^{\mathrm{H}}$ should be in the form of expectations, and they can be further simplified with the assumption that the mean forecast error is zero [18].

### D. *Model Reformulation*

With linear decision rules and Cantelli's inequality, (15)-(28) can be reformulated into the MISOCP model. At first three levels of covariance matrix are defined:

$$\boldsymbol{\Sigma}_{s,t}\in\mathbb{R}_{+}^{2\times2} \quad (41)$$

$$\boldsymbol{\Sigma}_{s,(t_1,t_2)}=\begin{bmatrix}\boldsymbol{\Sigma}_{s,t_1} & \boldsymbol{\Upsilon}_{s,(t_1,t_2)}\\ \boldsymbol{\Upsilon}_{s,(t_1,t_2)} & \boldsymbol{\Sigma}_{s,t_2}\end{bmatrix}\in\mathbb{R}^{4\times4} \quad (42)$$

$$\begin{aligned}\hat{\boldsymbol{\Sigma}}&=\begin{bmatrix}\boldsymbol{\Sigma}_{1,1} & \boldsymbol{\Upsilon}_{(1,1),(1,2)} & \cdots & \boldsymbol{\Upsilon}_{(1,S),(1,T)}\\ \boldsymbol{\Upsilon}_{(1,1),(1,2)} & \boldsymbol{\Sigma}_{1,2} & \cdots & \boldsymbol{\Upsilon}_{(1,S),(2,T)}\\ \vdots & \vdots & \ddots & \vdots\\ \boldsymbol{\Upsilon}_{(1,S),(1,T)} & \boldsymbol{\Upsilon}_{(1,S),(2,T)} & \cdots & \boldsymbol{\Sigma}_{S,T}\end{bmatrix}\in\mathbb{R}^{2ST\times2ST}\\ \boldsymbol{\Upsilon}_{(s_1,s_2),(t_1,t_2)}&=\mathbb{E}(\boldsymbol{\gamma}_{s_1,t_1}\boldsymbol{\gamma}_{s_2,t_2}{}^{\mathrm{T}})\end{aligned} \quad (43)$$

Here, (41) considers the spatial correlation of different renewable sources. (42) considers the spatial correlation and temporal correlation between two neighboring periods in the same scenario *s*, which is required to deal with ramping constraint (23). (43) considers the spatial correlation and temporal correlation between any two periods and any two scenarios that are required to deal with constraints such as (27) and (28). On this basis, the complete model reformulation of (15)-(28) is as follows:

$$\begin{aligned}&P_{i,y,s,t}^{\mathrm{WT,E}}+P_{i,y,s,t}^{\mathrm{WT,H}}=\sum_{\tau=1}^{y}P_{i,\tau}^{\mathrm{WT}}p_{i,s,t}^{\mathrm{WT}},\ \beta P_{i,y,s,t}^{\mathrm{WT,E}}+\beta P_{i,y,s,t}^{\mathrm{WT,H}}=\sum_{\tau=1}^{y}P_{i,\tau}^{\mathrm{WT}}\beta p_{i,s,t}^{\mathrm{WT}}\\&P_{i,y,s,t}^{\mathrm{PV,E}}+P_{i,y,s,t}^{\mathrm{PV,H}}=\sum_{\tau=1}^{y}P_{i,\tau}^{\mathrm{PV}}p_{i,s,t}^{\mathrm{PV}},\ \beta P_{i,y,s,t}^{\mathrm{PV,E}}+\beta P_{i,y,s,t}^{\mathrm{PV,H}}=\sum_{\tau=1}^{y}P_{i,\tau}^{\mathrm{PV}}\beta p_{i,s,t}^{\mathrm{PV}}\\&\beta p_{i,s,t}^{\mathrm{WT}}+\beta p_{i,s,t}^{\mathrm{PV}}=1\end{aligned} \quad (15')$$

$$\begin{aligned}f_{ij,y,s,t}^{\mathrm{HVDC}}&=P_{i,y,s,t}^{\mathrm{WT,E}}+P_{i,y,s,t}^{\mathrm{PV,E}}\\ \beta f_{ij,y,s,t}^{\mathrm{HVDC}}&=\beta P_{i,y,s,t}^{\mathrm{WT,E}}+\beta P_{i,y,s,t}^{\mathrm{PV,E}}\end{aligned} \quad (16')$$

$$\sqrt{(\beta f_{ij,y,s,t}^{\mathrm{HVDC}})^2\mathbf{1}^{\mathrm{T}}\boldsymbol{\Sigma}_{s,t}\mathbf{1}}\le\sqrt{\frac{\varepsilon}{1-\varepsilon}}\left(\sum_{\tau=1}^{y}\sum_{l=1}^{L_{ij,y}^{\mathrm{HVDC}}}\sigma_{ij,y,l}f_{ij}^{\mathrm{HVDC,max}}-f_{ij,y,s,t}^{\mathrm{HVDC}}\right) \quad (17')$$

$$\begin{aligned}&\sqrt{\left(\sum_{\mathrm{U}}\beta m_{i,y,s,t}^{\mathrm{D,U}}\right)^2\mathbf{1}^{\mathrm{T}}\boldsymbol{\Sigma}_{s,t}\mathbf{1}}\\&\le\sqrt{\frac{\varepsilon}{1-\varepsilon}}\left(\sum_{\tau=1}^{y}\sum_{l=1}^{L_{ij,y}^{\mathrm{HP}}}\sigma_{ij,y,l}m_{ij}^{\mathrm{HP}}-\sum_{\mathrm{U}}m_{i,y,s,t}^{\mathrm{D,U}}\right)\end{aligned} \quad (18')$$

$$\begin{aligned}m_{ij,y,s,t+1}^{\mathrm{HP}}&=m_{ij,y,s,t}^{\mathrm{HP}}+\sum_{\mathrm{U}}m_{i,y,s,t}^{\mathrm{D,U}}-\sum_{\mathrm{U}}m_{j,y,s,t}^{\mathrm{D,U}}\\ \beta m_{ij,y,s,t+1}^{\mathrm{HP}}&=\beta m_{ij,y,s,t}^{\mathrm{HP}}+\sum_{\mathrm{U}}\beta m_{i,y,s,t}^{\mathrm{D,U}}-\sum_{\mathrm{U}}\beta m_{j,y,s,t}^{\mathrm{D,U}}\end{aligned} \quad (19')$$

$$\sqrt{(\beta m_{ij,y,s,t}^{\mathrm{HP}})^2\mathbf{1}^{\mathrm{T}}\boldsymbol{\Sigma}_{s,t}\mathbf{1}}\le\sqrt{\frac{\varepsilon}{1-\varepsilon}}\left(\sum_{\tau=1}^{y}\sum_{l=1}^{L_{ij,y}^{\mathrm{HP}}}\sigma_{ij,y,l}m_{ij}^{\mathrm{HP,max}}-m_{ij,y,s,t}^{\mathrm{HP}}\right) \quad (20')$$

$$P_{i,y,s,t}^{\mathrm{WT,H}} + P_{i,y,s,t}^{\mathrm{PV,H}} = \sum_{\mathrm{M}} P_{i,y,s,t}^{\mathrm{M}}$$
$$\beta P_{i,y,s,t}^{\mathrm{WT,H}} + \beta P_{i,y,s,t}^{\mathrm{PV,H}} = \sum_{\mathrm{M}} \beta P_{i,y,s,t}^{\mathrm{M}} \tag{21'}$$

$$P_{i,y,s,t}^{\mathrm{M}} = x_{i,y,s,t}^{\mathrm{M}} P_{i,y,s,t}^{\mathrm{M,ON}} + u_{i,y,s,t}^{\mathrm{M}} P^{\mathrm{M,boot}}$$
$$\beta P_{i,y,s,t}^{\mathrm{M}} = x_{i,y,s,t}^{\mathrm{M}} \beta P_{i,y,s,t}^{\mathrm{M,ON}}$$
$$\sqrt{(\beta P_{i,y,s,t}^{\mathrm{M,ON}})^2 \mathbf{1}^{\mathrm{T}} \boldsymbol{\Sigma}_{s,t} \mathbf{1}} \le \sqrt{\frac{\varepsilon}{1-\varepsilon}} (P^{\mathrm{M,max}} - P_{i,y,s,t}^{\mathrm{M,ON}}) \tag{22'}$$
$$\sqrt{(\beta P_{i,y,s,t}^{\mathrm{M,ON}})^2 \mathbf{1}^{\mathrm{T}} \boldsymbol{\Sigma}_{s,t} \mathbf{1}} \le \sqrt{\frac{\varepsilon}{1-\varepsilon}} (P_{i,y,s,t}^{\mathrm{M,ON}} - P^{\mathrm{M,min}})$$

$$\sqrt{\begin{bmatrix} \beta P_{i,y,s,t+1}^{\mathrm{M}} \mathbf{1} \\ -\beta P_{i,y,s,t}^{\mathrm{M}} \mathbf{1} \end{bmatrix}^{\mathrm{T}} \boldsymbol{\Sigma}_{s,(t+1,t)} \begin{bmatrix} \beta P_{i,y,s,t+1}^{\mathrm{M}} \mathbf{1} \\ -\beta P_{i,y,s,t}^{\mathrm{M}} \mathbf{1} \end{bmatrix}}$$
$$\le \sqrt{\frac{\varepsilon}{1-\varepsilon}} (x_{i,y,s,t}^{\mathrm{M}} \Delta P^{\mathrm{M,max}} + (\chi_{i,y}^{\mathrm{M}} N^{\mathrm{M}} - x_{i,y,s,t}^{\mathrm{M}}) P^{\mathrm{M,max}} + P_{i,y,s,t}^{\mathrm{M}} - P_{i,y,s,t+1}^{\mathrm{M}}) \tag{23'}$$

$$m_{i,y,s,t}^{\mathrm{M}} = x_{i,y,s,t}^{\mathrm{M}} \left( b^{\mathrm{M}} (P_{i,y,s,t}^{\mathrm{M,ON}}) + c^{\mathrm{M}} \right)$$
$$\beta m_{i,y,s,t}^{\mathrm{M}} = x_{i,y,s,t}^{\mathrm{M}} b^{\mathrm{M}} (\beta P_{i,y,s,t}^{\mathrm{M,ON}}) \tag{24'}$$

$$f_{ij,y,s,t}^{\mathrm{HVDC}} = P_{j,y,s}^{\mathrm{D}} p_{j,s,t}^{\mathrm{D}}$$
$$\beta f_{ij,y,s,t}^{\mathrm{HVDC}} = \beta P_{j,y,s}^{\mathrm{D}} p_{j,s,t}^{\mathrm{D}} \tag{25'}$$

$$\sum_{\mathrm{U}} m_{i,y,s,t}^{\mathrm{S,U}} + m_{i,y,s,t}^{\mathrm{D,U}} = \sum_{\mathrm{M}} m_{i,y,s,t}^{\mathrm{M}}$$
$$\sum_{\mathrm{U}} \beta m_{i,y,s,t}^{\mathrm{S,U}} + \beta m_{i,y,s,t}^{\mathrm{D,U}} = \sum_{\mathrm{M}} \beta m_{i,y,s,t}^{\mathrm{M}} \tag{26'}$$

$$\sqrt{\begin{bmatrix} D_1 \beta m_{i,y,1,1}^{\mathrm{S,U}} \mathbf{1} \\ \vdots \\ D_S \beta m_{i,y,S,T}^{\mathrm{S,U}} \mathbf{1} \end{bmatrix}^{\mathrm{T}} \hat{\boldsymbol{\Sigma}} \begin{bmatrix} D_1 \beta m_{i,y,1,1}^{\mathrm{S,U}} \mathbf{1} \\ \vdots \\ D_S \beta m_{i,y,S,T}^{\mathrm{S,U}} \mathbf{1} \end{bmatrix}}$$
$$\le \sqrt{\frac{\varepsilon}{1-\varepsilon}} (m_{i,y}^{\mathrm{S,U}} - \sum_{s=1}^{S} D_s \sum_{t=1}^{T} m_{i,y,s,t}^{\mathrm{S,U}}) \tag{27'}$$

$$\sqrt{\begin{bmatrix} D_1 \beta m_{j,y,1,1}^{\mathrm{D,U}} \mathbf{1} \\ \vdots \\ D_S \beta m_{j,y,S,T}^{\mathrm{D,U}} \mathbf{1} \end{bmatrix}^{\mathrm{T}} \hat{\boldsymbol{\Sigma}} \begin{bmatrix} D_1 \beta m_{j,y,1,1}^{\mathrm{D,U}} \mathbf{1} \\ \vdots \\ D_S \beta m_{j,y,S,T}^{\mathrm{D,U}} \mathbf{1} \end{bmatrix}}$$
$$\le \sqrt{\frac{\varepsilon}{1-\varepsilon}} (m_{j,y}^{\mathrm{D,U}} - \sum_{s=1}^{S} D_s \sum_{t=1}^{T} m_{j,y,s,t}^{\mathrm{D,U}}) \tag{28'}$$